\documentclass[twocolumn,prl,aps,superscriptaddress,showpacs,amsmath,amssymb,floatfix,longbibliography]{revtex4-1}
\usepackage{graphicx}
\usepackage{amssymb}
\usepackage{amsmath}
\usepackage{epsfig}
\usepackage{color}
\usepackage{braket}
\usepackage[normalem]{ulem}

\definecolor{brown}{rgb}{0.3,0.1,0}

\begin{document}

\title{Exotic Spin Phases in Two Dimensional Spin-orbit Coupled Models: Importance of Quantum Fluctuation Effects}
\author{Chao Wang}
\author{Ming Gong}
\email{gongm@ustc.edu.cn}
\author{Yongjian Han}
\author{Guangcan Guo}
\author{Lixin He}
\email{helx@ustc.edu.cn}
\affiliation{CAS Key Laboratory of Quantum Information, Chinese Academy of Sciences, University of Science and Technology of China, Hefei, 230026, China}
\affiliation{Synergetic Innovation Center of Quantum Information
and Quantum Physics, University of Science and Technology of China, Hefei, 230026, China}
\begin{abstract}
We investigate the phase diagrams of the effective spin models derived from Fermi-Hubbard and Bose-Hubbard models with Rashba spin-orbit coupling,
using string bond states, one of the quantum tensor network states methods.
We focus on the role of quantum fluctuation effect in stabilizing the exotic spin phases in these models.
For boson systems, and when the ratio between inter-particle and intra-particle interaction $\lambda > 1$,
the out-of-plane  ferromagnetic (FM) and antiferromagnetic (AFM) phases obtained from quantum simulations are the same to those obtained from classic model.
However, the quantum order-by-disorder effect reduces the classical in-plane XY-FM and XY-vortex phases to the quantum X/Y-FM and X/Y-stripe phase
when $\lambda < 1$. The spiral phase and skyrmion phase can be realized in the presence of quantum fluctuation.
For the Fermi-Hubbard model, the quantum fluctuation energies are always important in the whole parameter regime.
%, which are much larger than those of the Bose model.
A general picture to understand the phase diagrams from symmetry point of view is also presented.
\end{abstract}


\maketitle

The ultracold atoms in optical lattice\cite{McKay11, Bloch08, Greiner02, Jaksch98} provide an excellent toolbox for simulating various spin models, such as
Heisenberg \cite{kuklov2003counterflow} model and Kitaev model\cite{Duan03} etc.,
and has been one of the central concepts in quantum simulations.
Along this line some primary results have been obtained\cite{Greif13, kim2010quantum, simon2011quantum}. The simplest ferromagnetic (FM) or
antiferromatic (AFM) Heisenberg spin models can be obtained in the deep Mott phase regime\cite{Greiner02} when the Hubbard model possesses rotational
symmetry.
The recent interest in the searching of exotic spin structures in optical lattice is stimulated by the experimental realization of spin-orbit coupling (SOC), which can
be regarded as the simplest non-Abelian gauge potential in nature\cite{Liu09,Spielman09,Zhai10,Lin11,Zhang11,Wu11,Dalibard11,Spielman11,Lawrence12,Zhai12,
Yun12,Spielman13,Gong13,Gong14,Spielman15,Li16,ChenShuai16,Huang16}.
In these cases, the effective spin models may become more complicated due to the appearance of some exotic terms, e.g., the
Dzyaloshinskii-Moriya (DM) \cite{DM58,DM60} interactions and their deformations.

The DM interaction has already been widely investigated
in solid materials\cite{Dagotto06, Cao09,Muhlbauer09,Masahito10,Yoshinori10,Yu10,Stefan11,Seki12,Nagaosa13,Wilson14}
and now it is resurfaced in ultracold atoms due to its
flexibility in experiments, e.g., the SOC interactions can be made much stronger than their counterpart in real materials.
Results based on classical simulations\cite{Radic12,Cole12,Gong15}, Ginzburg-Landau theory\cite{Roszler06,
Rowland16}, dynamical mean-field theory\cite{he2015bose} and spin wave expansion\cite{Ye15,Ye16,Ye16-2} have unveiled rich phase structures including spin spirals, skyrmions in the
presence of the frustrated interactions caused by the SOC: there are strong competition between spin-independent tunneling and the SOC induced
spin-flipping tunneling. However, the role of quantum fluctuation effect to the quantum phase diagrams in these models have not been thoroughly
investigated. Whether and how these phases can survive in the presence of quantum fluctuation are still unclear.

In this Letter, we investigate the quantum phase diagrams of the effective spin models with Rashba SOC, derived from Bose-Hubbard (BH)
model and Fermi-Hubbard (FH) model on a 12$\times$12 square lattice, using
recently developed string bond states, one of the tensor network states (TNS) methods\cite{Vidal03,Verstraete04,sandvik07,schuch08,Song14}.
The TNS methods provide promising tools to investigate quantum systems with frustrated interactions.
Details of the calculations are presented in Supplementary materials \cite{supp}.
We find whereas in some parameters regions the classic spin model can give qualitatively correct
ground states, in some regions, the quantum effects are crucial to get correct ground states.
 In particular for the fermion systems, the quantum effects are always important.

{\it Effective Spin Models}. For a BH model with Rashba SOC, the Hamiltonian can be written as
$H_\text{BH}=\mathcal{H}_0+\frac{U}{2} \sum_{i,\sigma}{n_{i\sigma}(n_{i\sigma}-1)}+\lambda U\sum_i{n_{i\uparrow}n_{i\downarrow}}$,
where $U$ and $\lambda U$ are on-site intra-particle and inter-particle interactions and $\mathcal{H}_0 =-t\sum_{\langle i,j\rangle}\Psi^\dagger_i\exp[-i\alpha {\bf e}_z\cdot(\vec{\sigma}\times {\bf e}_{ij})]\Psi_j$. Here $\Psi^\dagger=(b_{i\uparrow}^\dagger,
b_{i\downarrow}^\dagger)$, with $b_{i\sigma}^\dagger$ being the creation operator with site $i$ and spin $\sigma = {\uparrow, \downarrow}$ and ${\bf e}_{ij}$ being the
unit vector from site $i$ to $j$.
In the first Mott lobe ($U\gg t$), each site contains only one particle,
the effective spin model can be written as,
\begin{eqnarray}
    && H=J\sum_{\langle i,j\rangle_x}[\frac{\cos(2\alpha)}{\lambda}S_i^xS_j^x+\frac 1{\lambda}S_i^yS_j^y+\frac{\cos(2\alpha)}{\lambda}(2\lambda-1) S_i^z S_j^z \nonumber \\
    &&-\sin(2\alpha)(S_i^xS^z_j-S_i^zS_j^x)]+J\sum_{\langle i,j\rangle_y}[\frac 1{\lambda}S_i^xS_j^x + \frac{\cos(2\alpha)}{\lambda}S_i^yS_j^y \nonumber \\
    &&+\frac{\cos(2\alpha)}{\lambda}(2\lambda-1)S_i^zS_j^z - \sin(2\alpha)(S_i^yS^z_j-S_i^zS_j^y)], \label{eq-HJ}
\end{eqnarray}
where $J$=$-4t^2/U$$<$0, and
$\langle i,j\rangle_{\mu}$ means the nearest neighbors in the $\mu=x, y$ directions.
In this model $\alpha$ determines the strength of SOC, and $\lambda$ represents the anisotropy of the exchange interactions.
Similarly in the FH  model, the Hamiltonian reads as
$H_\text{FH}= \mathcal{H}_0+\sum_i Un_{i\uparrow}n_{i\downarrow}$ where $\mathcal{H}_0$ has the same form as boson model with $\Psi^\dagger$ replaced by $(f_{i\uparrow}^\dagger, f_{i\downarrow}^\dagger)$,
where $f_{i\sigma}^\dagger$ is the fermion creation operator at site $i$ and spin $\sigma = {\uparrow, \downarrow}$. The corresponding effective spin model equals to that in Eq.\ref{eq-HJ} at $\lambda = 1$ except that now $J = 4t^2/U >$0 due to Pauli exclusion principle. Hereafter we let $4 t^2/ U = 1$ for convenience.

\begin{figure}[t]
		\centering
		\includegraphics[width=0.45\textwidth]{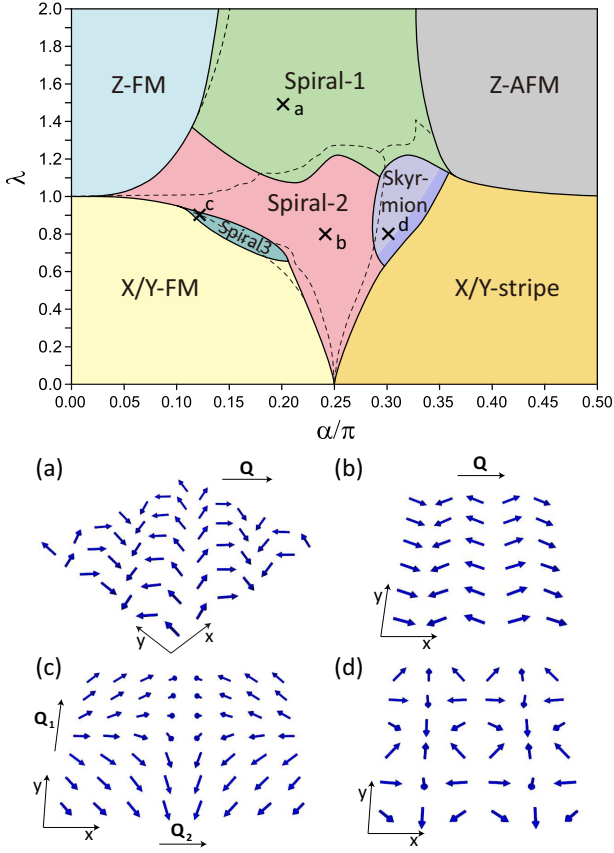}
                \caption{(Color online). Upper panel: Phase diagram for Eq. \ref{eq-HJ}, calculated from TNS method.  Z-FM (Z-AFM) denote the ferromagnetic (antiferromagnetic) phase with spin
                polarized along the $z$ direction, and X/Y means that the spins are polarized along either $x$ or $y$ direction. For a comparison the phase boundaries determined by classical simulations are shown in dashed lines.
                Lower panel: Spin textures of (a) Spiral-1 phase with spins spiral along ${\bf Q}={\bf e}_x-{\bf e}_y$ direction; (b) Spiral-2 phase with spins spiral along ${\bf Q}={\bf e}_x$ direction; (c) Spiral-3 phase spins spiral along both ${\bf Q}_1={\bf e}_y$ and ${\bf Q}_2={\bf e}_x$ directions; (d) 3$\times$3 skyrmion phase.}
    \label{fig-fig1}
\end{figure}

The following order parameters are used to distinguish different phases. Firstly, the static magnetic structure factor is defined as
[$\mu =x, y, z$, $i = (i_x, i_y)$],
\begin{equation}
    S^\mu(\textbf{k})=\frac 4{L^2}\sum_{i,j}\langle  S^\mu_{i}\cdot  S^\mu_{j}\rangle e^{i[(i_x-j_x)k_x+(i_y-j_y)k_y]/L}.
    \label{eq-sf}
\end{equation}
on a $L\times L$ square lattice. For the FM and AFM phases along $\mu$-direction, $S^\mu(\textbf{k})$ has peaks at ${\bf k} = (0, 0)$ and
$(\pi, \pi)$, respectively;
and in the strip phase the strongest peaks happen at ${\bf k} = (0, \pi)$ or $(\pi, 0)$.
We also define the spiral and skyrmion order parameters in real space as\cite{Cole15},
%should be distinguished in combination of the following correlators by monitoring the spin configuration,
\begin{equation}
        \text{Sp}_\mu(i,j)=16 \langle {\theta}_\mu^{i} {\theta}_\mu^{j}\rangle, \quad \text{Sk}(i,j)= 64 \langle v_s^{i} v_s^{j}\rangle,
        \label{eq-sp_old}
    \end{equation}
where $\theta_x^i= ({\bf S}_{i}\times {\bf S}_{i+{\bf e}_x})_y$ and $\theta_y^i= ({\bf S}_{i}\times {\bf S}_{i+{\bf e}_y})_x$ are related to the relative planer spin angles for
spins at site $i$ and $i+{\bf e}_{\mu}$. To account for the three dimensional spin alignment effect,
we define the spin volume constructed by the spins at the three neighboring sites
as $v_s^i= {\bf S}_{i}\cdot({\bf S}_{i+{\bf e}_x}\times {\bf S}_{i+{\bf e}_y})$.
In the co-planar spiral phase, $v_s^i = 0$ exactly, but it is nonzero in the skyrmion phases.  %Notice
%that while these two new correlators exactly equal to zero for the FM, AFM and stripe phase in the classical picture,
%they may become finite in the quantum treatment when $|i-j|$ is small due to strong fluctuation effect in short range. Thus
To determine the long-range order of the system, we calculate the order parameters as
\begin{equation}
    \text{Sp}_\mu= \sum_i {1 \over L^2} |\text{Sp}_\mu(i,i+l)|, \text{Sk}= \sum_i {1 \over L^2} |\text{Sk}(i,i+l)|,
    \label{eq-sp}
\end{equation}
where $l = (L/2, L/2)$ to make $|i-j|$ as large as possible and $i$ is averaged over the whole lattice for better numerical accuracy.

	\begin{figure}[htb]
		\centering
		\includegraphics[width=0.42\textwidth]{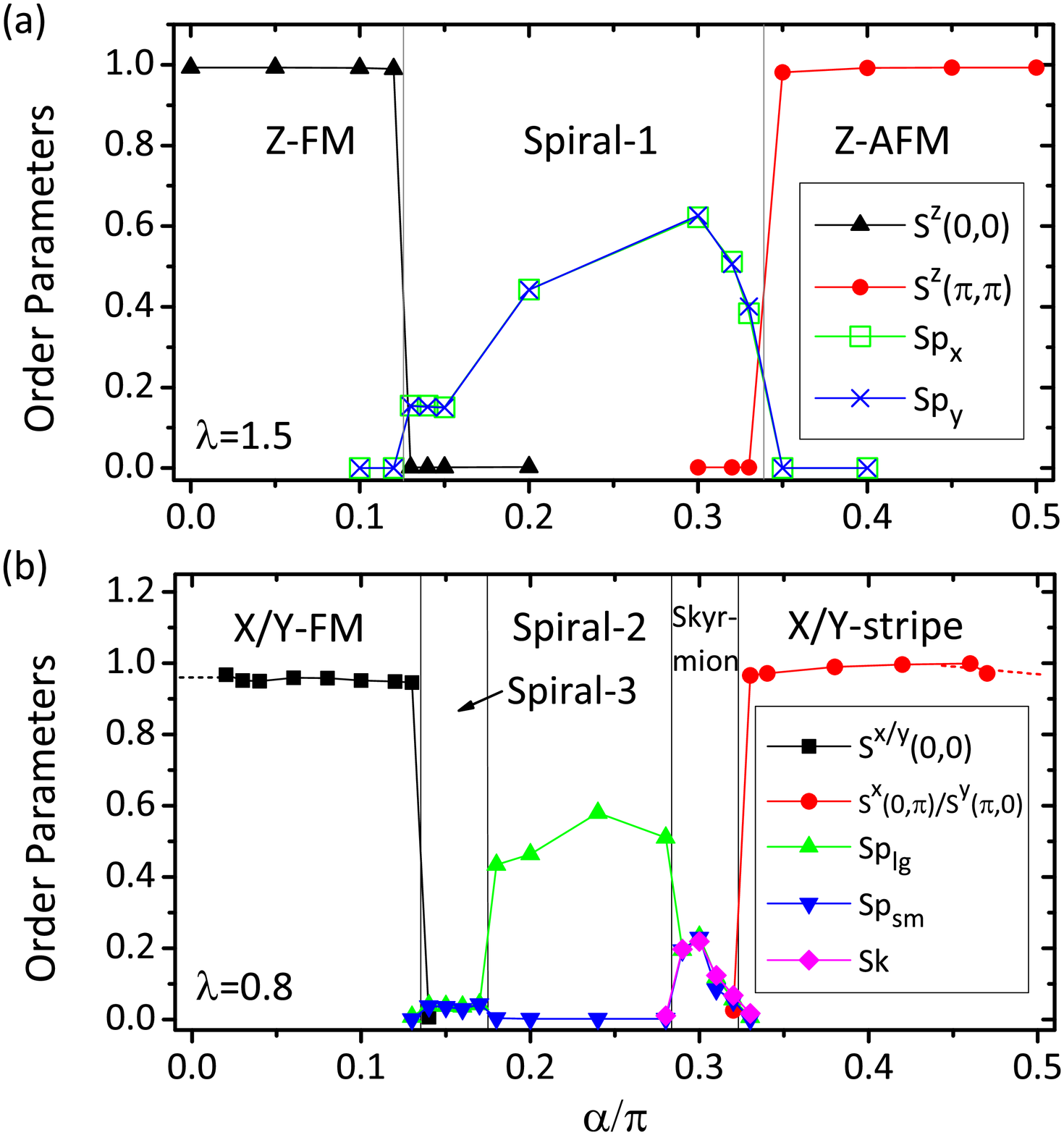}
                \caption{(Color online). Order parameters for the effective spin model derived from Bose-Hubbard model at
                (a) $\lambda=1.5$; and (b)$\lambda=0.8$, where $ \text{Sp}_{\rm lg}$ [$ \text{Sp}_{\rm sm}$]
                is the larger (smaller) one of $\text{Sp}_{x}$ and $\text{Sp}_{y}$  (see text for details). }
     %            where the spiral-1 phase is characterized by $\langle \text{Sp}_x(i) \rangle=\langle \text{Sp}_y (i)\rangle\neq 0$.
	%	(b) $\lambda=0.8$, where in the spiral-2 phase, the larger one of (denoted by is non-vanishing, and the smaller one (denoted by) vanishes. However, both of them are non-vanishing in the spiral-3 phase.}
		\label{fig-fig2}
	\end{figure}

{\it Phase Diagram for Boson.} The phase diagram is presented in Fig. \ref{fig-fig1} and corresponding order parameters are given in Fig. \ref{fig-fig2}
at $\lambda$=1.5 and $\lambda$=0.8 and $\alpha \in [0, \pi/2]$.
The spin model in Eq. \ref{eq-HJ} possesses some unique symmetries, which is crucial
to understand this phase diagram. Firstly, the Hamiltonian in Eq. \ref{eq-HJ} is invariant upon operation $\alpha$$\rightarrow$$\pi-\alpha$ and $S_i^{x,y}\rightarrow -S_i^{x,y}$, which is equivalent to
the transformation $U^\dagger_\uparrow b_{i\uparrow} U_{\uparrow} = -b_{i\uparrow}$, where $U_{\uparrow} = \exp(i\pi \sum_i n_{i\uparrow})$
in the original BH model. This symmetry directly leads to $U^\dagger_\uparrow H(\alpha, \lambda) U_\uparrow = H(\pi -\alpha, \lambda)$, i.e.,  the phase
diagram should be symmetric about $\alpha = {\pi \over 2}$. Therefore we only  show the result for $\alpha \in [0, \pi/2]$.

We first discuss the phase diagram at four corners, where $\alpha\sim$ 0 or $\pi/2$ and $\lambda \ll 1$ or $\lambda \gg 1$.
When $\alpha =0$, i.e., in the absence of SOC, the original spin model can be reduced to
an effective XXZ spin model, with $J_x= J_y = -{1 / \lambda}$, and
$J_z= -{(2\lambda-1)/\lambda}$.
When $\lambda > 1$, $|J_z| >|J_x|$, the ground state is a Z-FM state, i.e., all spins are ferromagneticlly aligned along the $z$ direction.
Our TNS calculations show that for small $\alpha \lesssim$0.15, the ground state is still Z-FM,
as determined by the order parameters shown in Fig. \ref{fig-fig2}a.
In this region, the quantum simulations yield the same ground state as the classic one, suggesting the minor role of quantum fluctuation effect.

Interestingly, at $\alpha = {\pi \over 2}$, the model can be mapped to
the $\alpha$=0 case via a symmetry transformation, $\mathcal{U} = \prod_{i} e^{-i{\pi\over 2} i_x \sigma_x} e^{-i{\pi\over 2} i_y\sigma_y}$,
i.e., $\mathcal{U}^\dagger H_{\pi\over 2} \mathcal{U} = H_{0}$.
%This transition corresponds to the particle-hole transformation $c_{i\sigma} \rightarrow c_{i\sigma}^\dagger$ in the BH model.
Use this transformation, we immediately see that the ground state near $\alpha = {\pi \over 2}$ is a Z-AFM.
We therefore see that these two limits ($\alpha$=0 and  $\alpha$=${\pi \over 2}$) should have the exact same energies,
and the quantum effects are small in both phases, which are confirmed by the numerical results.

However, there are dramatic difference in the case of $\lambda <$1 where the in-plane exchange energy dominates.
The order parameters calculated by TNS at $\lambda$=0.8 are shown in Fig. \ref{fig-fig2}.
In the region of 0 $< \alpha/\pi <$ 0.13, the ground state is a FM phase,
with all spins are polarized along either $x$ or
$y$ direction, which we denote as X/Y-FM phase.
Remarkably this phase is very different from what is obtained from the
classical spin model, which gives a rotational invariant
FM state~\cite{Cole12} with all spins lay in the $x$-$y$ plane (dubbed as XY-FM).
To understand this difference, we note that the in-plane rotational symmetry is not inherent
of the original Hamiltonian, which possesses only $C_4$ symmetry.
The rotational invariance of the ground state in the classic model is due to the accidental degeneracy
because the ground state of classic model happen to has $S_z$=0.
When quantum fluctuation is introduced, it breaks the accidental degeneracy
and restore the $C_4$ symmetry of the original Hamiltonian, which therefore
single out a ground state with lower energy than the classical solution.
This is the known as order-by-disorder mechanism\cite{villain80,shender82}.
Again, we can apply symmetry transformation $\mathcal{U} = \prod_{i} e^{-i{\pi\over 2} i_x\sigma_x} e^{-i{\pi\over 2} i_y\sigma_y}$ near  $\alpha=\pi/2$,
which yields a X/Y-stripe phase (as confirmed by numerical results) for quantum spin model,
in contrast to the 2$\times$2 vortex state obtained from classical simulations.

\begin{figure}
	\centering
	\includegraphics[width=0.46\textwidth]{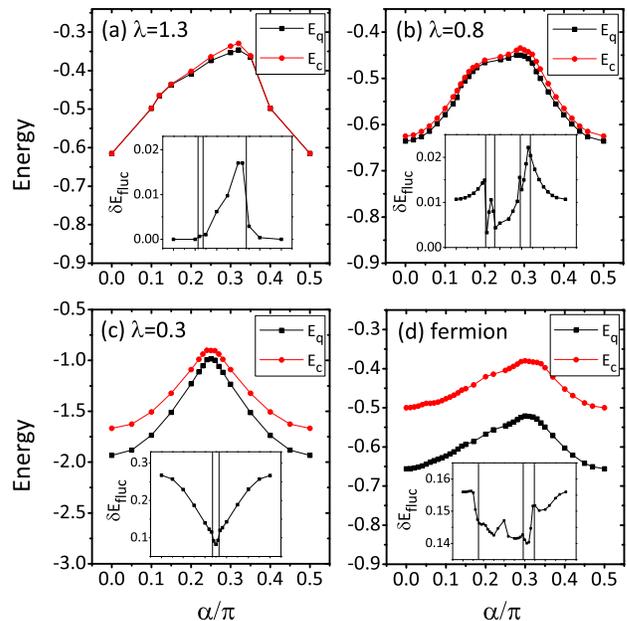}
        \caption{(Color online). Ground state energies from the classical ($E_\text{c}$) and quantum ($E_\text{q}$) simulations
         at (a) $\lambda = 1.3$, (b) $\lambda = 0.8$ and (c) $\lambda = 0.3$ for boson, and (d) for fermion models.
         Insets give the corresponding  fluctuation energy
$\delta E_{\rm fluc}$. The vertical lines are the phase boundaries calculated by TNS.}
        \label{fig-fig3}
\end{figure}

The line $\lambda = 0$ in principle can not he achieved due to the energy-costless double occupation. However this limit can still be defined
in the sense of $\lim_{\lambda \rightarrow 0} \lambda H_{\lambda}= - \sum_{\langle i,j\rangle _x} [\cos(2\alpha) (S_i^x S_j^x+S_i^z S_j^z)+S_i^y S_j^y]
-\sum_{\langle i,j\rangle _y} [\cos(2\alpha)(S_i^y S_j^y +S_i^z S_j^z)+S_i^x S_j^x]$.
Obviously when $\alpha = {\pi \over 4}$,
	\begin{equation}
            \lim_{\lambda \rightarrow 0} \lambda H_{\lambda,\alpha = {\pi \over 4}}= -(\sum_{\langle i,j\rangle_x}S_i^y S_j^y+\sum_{\langle i,j\rangle_y} S_i^x S_j^x)
	\end{equation}	
which gives a compass model due to the strong coupling between the spins and directions\cite{Nussinov15}.
This model can not be solved exactly; however it can be shown exactly that the ground state is $2^{L+1}$-fold degenerated for a $L \times L$ square
lattice\cite{Dorier05,You10,Brzezicki13}. It therefore corresponds to a critical boundary between the X/Y-FM and X/Y stripe phases since any deviation from
this critical point by varying the parameters ($\lambda$ and $\alpha$) can break the degeneracy and open an energy gap.
The classical and quantum simulations yield the same critical point.

We next try to understand the spiral and skyrmion phases in the presence of strong DM
interaction. The order parameters are shown in Eq. \ref{eq-sp} (and the corresponding spin textures are shown in the lower panel of Fig. \ref{fig-fig1}).
The spiral-1 phase has two degenerate states spiral along either ${\bf e}_x+{\bf e}_y$ or ${\bf e}_x-{\bf e}_y$ direction.
For these two cases the strongest peaks in the structure factor $S({\bf k})$ appear at ${\bf k} = \pm(k_0,k_0)$ and
${\bf k} = \pm(k_0,-k_0)$, respectively, where $k_0$ can be smoothly tuned by $\alpha$ and $\lambda$. However, due to
the finite size used in the simulation, only $k_0$=$\frac{2\pi}3$, $\frac{\pi}2$, $\frac{\pi}3$ and $\frac{\pi}6$ are observed,
which are commensurate with the system size.
In this phase, the skyrmion order $ \text{Sk} \sim 0$, whereas $ \text{Sp}_{x} =  \text{Sp}_y \ne 0 $ are strongest among all the order parameters.
The spiral-2 phase has two degenerate states, one is a spin spiral along $x$ direction, and other one is along $y$ direction.
Therefore, only one of the order parameters, either
$\text{Sp}_x$ or $ \text{Sp}_y$ (see Fig. \ref{fig-fig2}b) is nonzero.
In contrast, in spiral-3 phase, $ \text{Sp}_x$=$ \text{Sp}_y$, both are nonzero.
Spiral-3 phase is also observed in the classical model, and compared to the classical model,
the spiral-3 phase region is greatly suppressed in the quantum model.
In the skyrmion phase, the structure factor exhibits strongest peaks at
${\bf k}$=$(\pm k_0, 0)$ and $(0, \pm k_0)$.
Furthermore the non-conplaner of spin alinement induce a finite srkymion order $\text{Sk}$.
The skyrmion phase is Neel type\cite{kezsmarki2015neel} and has a period $3\times 3$
(light purple region in Fig. \ref{fig-fig1}) or larger(dark purple region in Fig. \ref{fig-fig1}),
which is consistent with the numerical results for the classic spin model \cite{Cole12}.

To understand the quantum effects in a more quantitative way, we plot the ground state energies per site for
$\lambda$=1.3, 0.8, 0.3 in Fig. \ref{fig-fig3} a - c respectively obtained from classical simulations ($E_{\rm c}$)
and full quantum mechanical TNS simulations ($E_{\rm q}$).
In the inserts, we also show the energy differences
\begin{equation}
    \delta E_\text{fluc} = E_\text{c} - E_\text{q} \, .
    \label{eq-deltaE}
\end{equation}
Obviously $E_\text{c} \ge E_\text{q}$, thus $\delta E_\text{fluc} \ge 0$.
From Fig. \ref{fig-fig3}a. we find that when $\alpha =0$,  and $\lambda$=1.3,
$E_\text{c} = -0.61537$, and $E_\text{q} = -0.61538$ in the 12$\times$12 lattice,
while the exact classical energy in a infinite size system is
$ E_\text{c}^{\infty}= {2\lambda -1\over 2\lambda} = -0.61538$.
This agreement can be understood using the Holstein-Primarkoff (HP) transformation to the XXZ model (see \cite{supp}) due to the disappearance
of pairing (or condensate) term, thus $\delta E_\text{fluc} = 0$ exactly. In fact the XXZ model can be used
as a benchmark for the TNS method, which shows great accuracy in this problem.
As shown in Fig. \ref{fig-fig3}a,
$\delta E_\text{fluc} \sim 0$ in the whole Z-FM and Z-AFM phase regimes, even when $\alpha \neq$0.
In the spiral phase, $\delta E_\text{fluc} \sim$ 0.01 - 0.02 is more significant.

The fluctuation energy increases with the decreasing of $\lambda$. At $\lambda$=0.8, $\delta E_\text{fluc} \sim$0.01 in the X/Y-FM and X/Y strip
phases, which is about 4\% of the total energies.
However, even though this energy difference seems not very large,
the ground states predicted by classical model and quantum model are totally different.
Full quantum treatments are therefore required to capture the correct physics in these phases.
$\delta E_\text{fluc}$ is also different for different phases, which is most significant in the skyrmion phase, where $\delta E_\text{fluc}\sim$0.02.
When $\lambda$ further decrease to 0.3, $\delta E_\text{fluc}$ increases dramatically. It is about 0.1 - 0.3 in the X/Y-FM and X/Y strip phases, which
counts almost 10\% - 20\% of the total energies. The strong quantum fluctuation at small $\lambda$ suppresses the
spiral-3 and skyrmion phases compared to the the classical phase diagram (see Fig.~\ref{fig-fig1}).

\begin{figure}[t]
\centering
\includegraphics[width=0.48\textwidth]{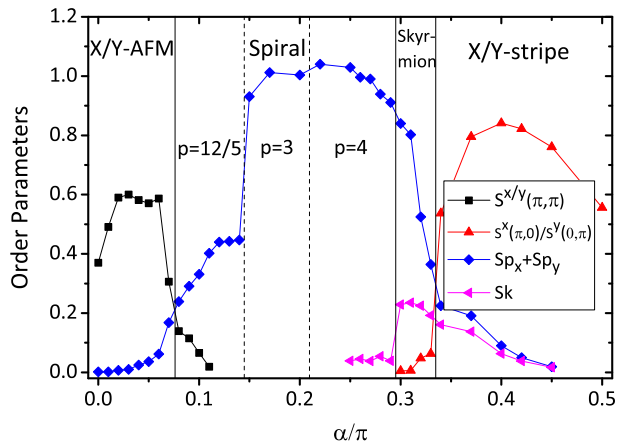}
    \caption{(Color online). Order parameters for the effective spin model derived from Fermi-Hubbard model.
    }
\label{fig-fig4}
\end{figure}

{\it Phase Diagrams for Fermion.} For the spin model from FH model, we have $J = 4t^2/U>$0, and $\lambda$=1.
Therefore $\alpha$ serves as the only adjustable parameter in this model.
The calculated phase diagram and the order parameters from the TNS method are presented in
Fig. \ref{fig-fig4}. Similar to the phase diagrams in the bosonic system, we find  X/Y-AFM phase when $\alpha/\pi < 0.08$ and X/Y-Stripe phase when
$\alpha/\pi \in [0.34, 0.5]$ (the mirror symmetry about $\alpha = {\pi \over 2}$ is assumed). As before the classical model predicts a rotational
invariant AFM and vortex phases in the $x$-$y$ plane, which reduce to the X/Y-AFM and X/Y-stripe phase due to order-by-order effect.
Between the AFM and stripe phases, there are spiral phases and one skyrmion phase. The spiral phase may also be distinguished by the period
$p = {12\over 5}, 3, 4$ which can be accommodated by our simulation sizes. In these phases the skyrmion order almost equal to zero and the spiral
order dominates. However, when $\alpha/\pi \in [0.29,0.34]$ , the skyrmion order become important although the spiral order is still nonzero,
similar to that in Fig. \ref{fig-fig2}b.

The quantum fluctuation energy is much more pronounced in the FH model
than in the BH model for all phases, as depicted in Fig.\ref{fig-fig3}d.
For $\alpha$= 0, we find $E_\text{c} = -0.5$, and $E_\text{q} = -0.6579$, thus
$\delta E_\text{fluc} = 0.1579$.
In the AFM and strip phases, $\delta E_\text{fluc}$ is about 30\% of the total energy.
The large quantum fluctuation energy in the AFM state is due to that there are vast Hilbert spaces
near the $S$=0 that are energetically close to the ground state.
The $\delta E_\text{fluc}$ is slightly small in the spiral phase and skyrmion phase, but still significant.

It is very interesting to note that the Z-AFM state in BH model however has very small $\delta E_\text{fluc}$,
in sharp contrast with the AFM state derived from the FH model. To understand this difference, we note that the Z-AFM state in BH model can be
mapped to the Z-FM state via symmetry transformations, which has small quantum fluctuation energy.
Therefore, even though the two AFM states appear very similar to each other at the {\it classical} level,
their physics are entirely different.
More fundamentally, this difference is rooted from the different statistic properties between bosons and fermions.

{\it Conclusion}. We address the role of quantum fluctuation effect on the possibilities on observing the exotic spin structures in the spin-orbit coupled BH and FH
models on a square lattice using TNS method. While for the out-of-plane FM and AFM phases the classical and quantum solution are the same, we find
that the quantum order-by-disorder effect reduces the classical in-plane XY-FM and XY-vortex phases to the quantum X/Y-FM and X/Y-stripe phase. Moreover, the spiral phase and skyrmion phase
can still be found even in the presence of quantum fluctuating effect. The structure of the phase diagrams are also understood from the symmetry point of view.

{\it Acknowledgement}.
This work was funded by the Chinese National Science Foundation No. 11374275, 11474267,
the National Key Research and Development Program of China No. 2016YFB0201202.
M.G. acknowledges the support by the National Youth Thousand Talents Program No. KJ2030000001,
the USTC start-up funding No. KY2030000053 and the CUHK RGC Grant No. 401113.
The numerical calculations have been done on the USTC HPC facilities.
M.G. Thank W. L. You for valuable discussion about compass model.

%\bibliography{refs,supp}


\end{document}